\documentclass[10pt,a4paper]{article}
    \usepackage[T1]{fontenc}
    \usepackage{inputenc}
    \usepackage[english]{babel}
    \usepackage[nohead,includefoot,margin=2cm]{geometry}
    \usepackage[compact,raggedright,small]{titlesec}
    \usepackage[affil-it]{authblk}
    \usepackage[runin]{abstract}
    \usepackage{multicol}
    \usepackage{blindtext}
\usepackage{subfig}
\usepackage{graphicx}
    \newcommand*{\pacsname}{PACS numbers}

    \addto{\Affilfont}{\small}

    \title{\large\bfseries Analog Optical Computing by Half-Wavelength Slabs}

    \author[1]{Farzad Zangeneh-Nejad}
    \author[1,*]{Amin Khavasi}
    \author[1]{Behzad Rejaei}
    \affil[1]{Department of Electrical Engineering, Sharif University of Technology, Tehran, Iran, P.O. Box 11555-4363}
\affil[*]{Corresponding author: khavasi@sharif.edu}
    
\begin{document}

       \maketitle

       \begin{abstract}
A new approach to perform analog optical differentiation is presented using half-wavelength slabs. First, a half-wavelength dielectric slab is used to design a first order differentiator. The latter works properly for both major polarizations, in contrast to designs based on Brewster effect [Opt. Lett. 41, 3467 (2016)]. Inspired by the proposed dielectric differentiator, and by exploiting the unique features of graphene, we further design and demonstrate a reconfigurable and highly miniaturized differentiator using a half-wavelength plasmonic graphene film. To the best of our knowledge, our proposed graphene-based differentiator is even smaller than the most compact differentiator presented so far [Opt. Lett. 40, 5239 (2015)].
       \end{abstract}
\indent Despite the prevalence of digital computers, analog computing is still of great significance since it can overcome certain inherent limitations of digital computation such as data conversion loss. For instance, analog computation is presented as an important theoretical frame work for discussing computation in many natural systems \cite{1}-\cite{4}.These advantages and applications, together with the recent advances in optical technology, has led to the emergence of analog optical computing as a new concept \cite{4}-\cite{e9}.\\
\indent The proposed approaches to perform analog optical computing can be fundamentally divided into two major classes, according to whether the computations are performed in temporal or spatial domain. Devices based on time domain calculations \cite{7}-\cite{e13}, however, are relatively large, making spatial computation more attractive.\\
\indent To perform spatial analog computations two different methods have been investigated: metasurface approach and Green's function(GF) approach. The GF method has attracted more attention due to less complex fabrication and ease of miniaturization \cite{10}. Here, mathematical computation is carried out using a multilayered slab which is homogeneous along two directions, and the operator of choice is realized by processing the optical field as it travels through the individual layers of the structure.\\
\indent Yet, there were two major weaknesses restricting the performance of the GF approach. First, due to the reflection symmetry of such systems, it was until recently not feasible to realize the associated Green's functions of many important operators having an odd symmetry in the spatial Fourier domain such as the operator of first-order differentiation \cite{10}. Second, the simplex optimization method used to calculate values of relative permittivities and thicknesses of slabs led to non-realistic values reducing the practical value of such structures \cite{10}.\\
\indent In a recent report \cite{11}, by breaking the reflection symmetry using an obliquely incident wave, and by exploiting Brewster effect, we demonstrated a first-order differentiator for the first time. This method circumvented the aforementioned drawback of GF method in realizing odd operators, and could be readily implemented. However, the design proposed still suffers from two major shortcomings. First, it is polarization-dependent since Brewster effect only occurs for incident waves with TM polarization. Second, it only works properly for a specific incident angle, i.e. the Brewster angle. \\
\indent In this paper, we aim to overcome these limitations by introducing a new approach based on a half-wavelength dielectric slab. We shall show that the resulting first order differentiator works appropriately for both TE and TM polarizations of the incident wave. Motivated by the performance of the proposed half-wavelength dielectric slab differentiator, and by exploiting the unique features of graphene in supporting plasmonic waves, we then design and demonstrate a differentiator using a planar half-wavelength plasmonic graphene film. It is worth noting that this device can be tuned by electrically adjusting the chemical potential of the graphene film through a gate. Moreover, because of the short wavelength of the plasmoic waves, the graphene-based differentiator is very compact. In fact, It is much smaller than the previously reported graphene-based analog optical differentiator in \cite{12}.\\
\begin{figure}
\centering
	\includegraphics[width=0.45\textwidth]{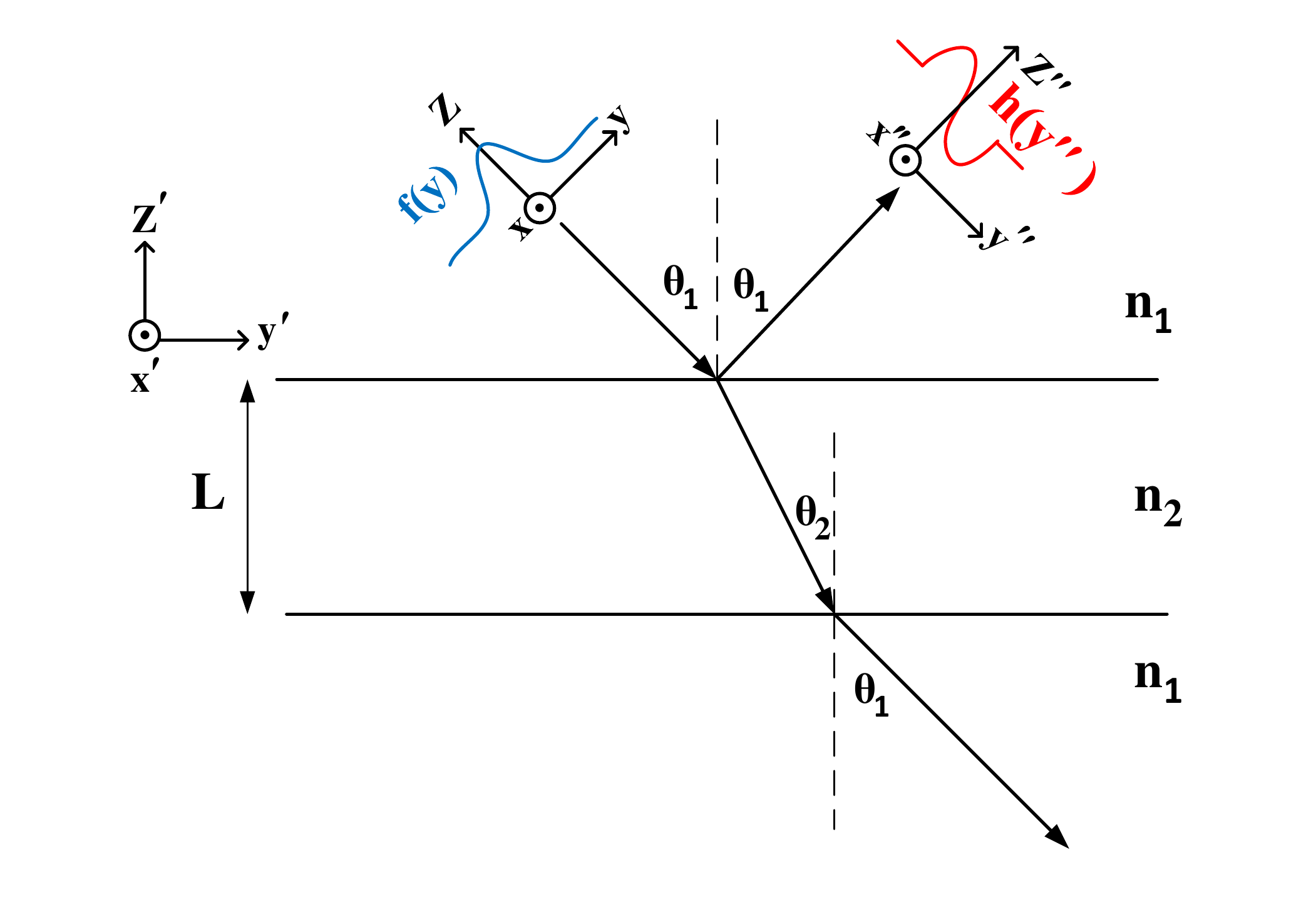}
	\captionsetup{justification=justified}
	\caption{Configuration of the proposed structure to perform the operator of first order derivative.}\label{fig1}
\end{figure}
\indent The schematic diagram of our first proposed structure is shown in Fig. 1, in which a dielectric slab is used to perform  first order differentiation. An arbitrary wave with the profile $f(y)$ is obliquely incident on a dielectric layer with length $L$ and refractive index of $n_{2}$, which is surrounded by two semi-infinite dielectric media having the refractive index of $n_{1}$. The angles of incidence and refraction are assumed to be $\theta_{1}$ and $\theta_{2}$, respectively. By suitably adjusting the parameter $L$, the reflection coefficient of the proposed structure vanishes for both TE and TM polarization. As outlined in \cite{11}, if the corresponding reflection coefficient is considered to be the Green's function of the structure, the reflected field $h(y^{\prime\prime})$ will be the first order derivative of $f(y)$, provided that the spatial bandwidth of $f(y)$ is small.\\
\indent Figs. 2(a) and (b) show the equivalent circuit models of the proposed differentiator for TE and TM polarizations. The semi-infinite dielectric media are modeled by the impedances  $Z=\frac{\omega\mu}{k_{0}n_{1}cos(\theta_{1})}$ and $Z=\frac{k_{0}cos(\theta_{1})}{\omega n_{1}}$ for the TE, respectively, TM polarizations. The dielectric layer is modeled as a transmission line with characteristic impedance of $Z_{0}=\frac{\omega\mu}{k_{0}n_{2}cos(\theta_{2})}$ for the TE case and $Z=\frac{k_{0}cos(\theta_{1})}{\omega n_{1}}$ for TM case, propagation constants of $\beta=k_{0}n_{2}cos(\theta_{2})$, and lengthes of $L$. Using the equivalent circuits shown, the reflection coefficient of the structure for both TE and TM polarizations can be expressed as
\begin{equation}\label{1}
R=\frac{Z_{in}-Z}{Z_{in}+Z}
\end{equation}
in which
\begin{equation}\label{2}
Z_{in}=Z_{0}\frac{Z+jZ_{0}tan(\beta L)}{Z_{0}+jZtan(\beta L)}
\end{equation}
\indent If $L$ is chosen such that $\beta L=\pi$, the reflection coefficient of the structure will be zero. In this case, the dielectric layer with the permittivity $n_{2}$ behaves just like a half wavelength transmission line transforming the load impedance $Z$ to the input side of the line. As a result, the input impedance of the line matches the impedance of the first semi-infinite dielectric medium, i.e. $Z_{in}=Z$, and consequently the reflection coefficient vanishes. As shown in \cite{11}, such a zero of the reflection coefficient can then be exploited to perform the operator of first order derivative.\\
\begin{figure}
\centering
     \subfloat[\label{subfig-1:dummy}]{%
       \includegraphics[width=0.45\textwidth]{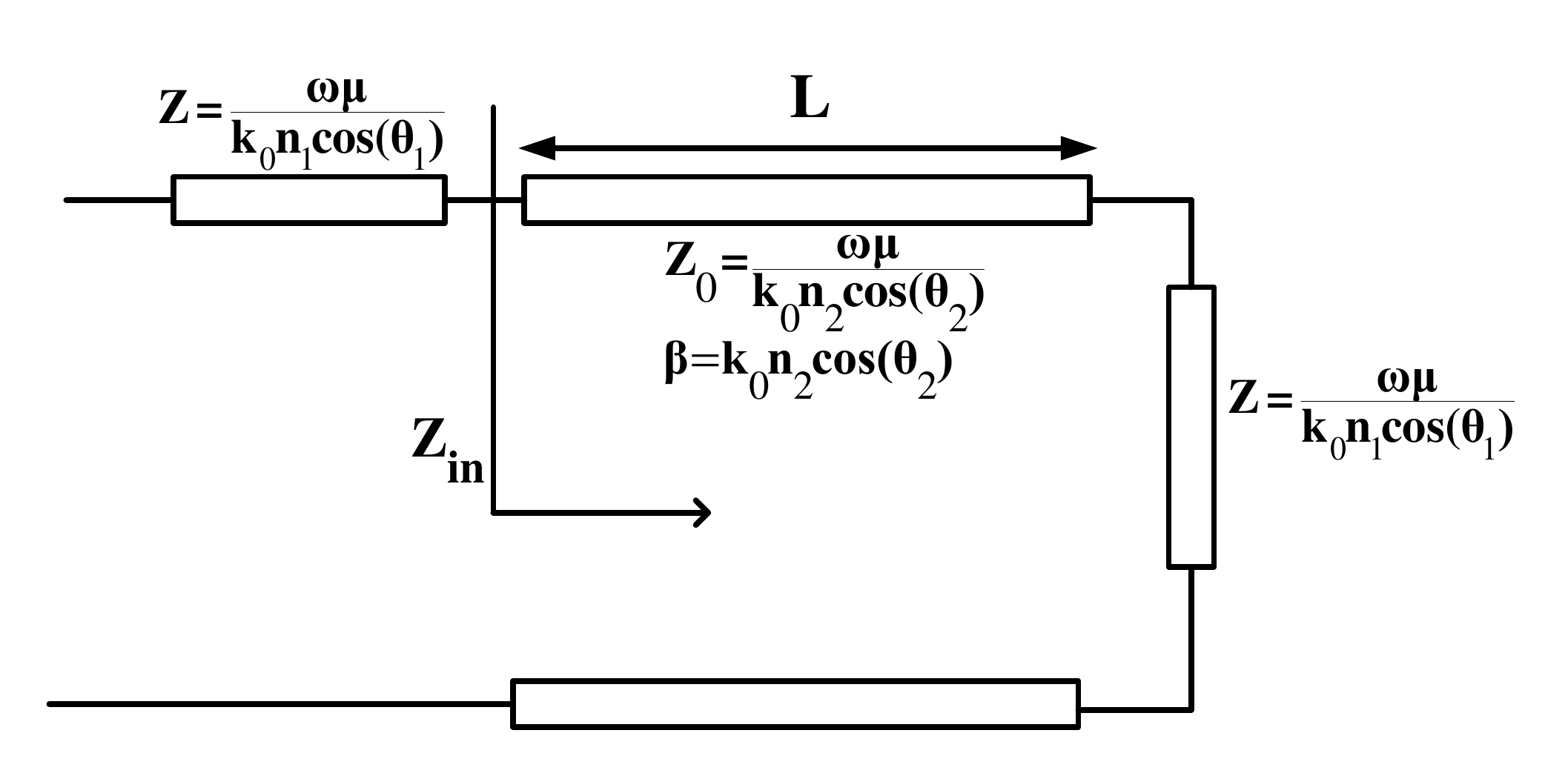}
     }
     \hfill
     \subfloat[\label{subfig-2:dummy}]{%
       \includegraphics[width=0.45\textwidth]{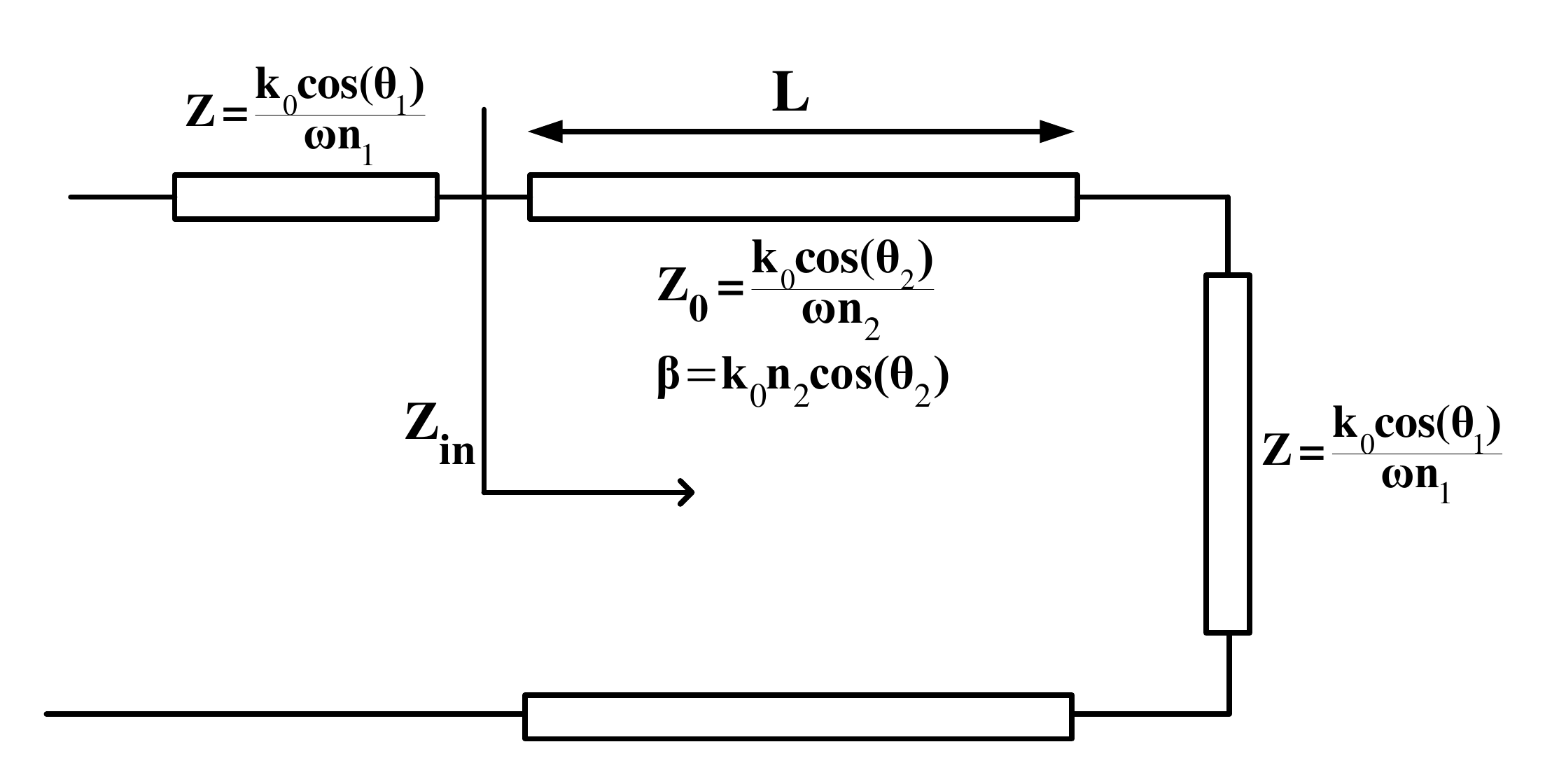}
     }
     \caption{Equivalent circuit model of presented structure for (a) TE and (b) TM polarization of incident wave.}
     \label{fig:dummy}
\end{figure}
\begin{figure}[!h]
\centering
     \subfloat[\label{subfig-1:dummy}]{%
       \includegraphics[width=0.25\textwidth]{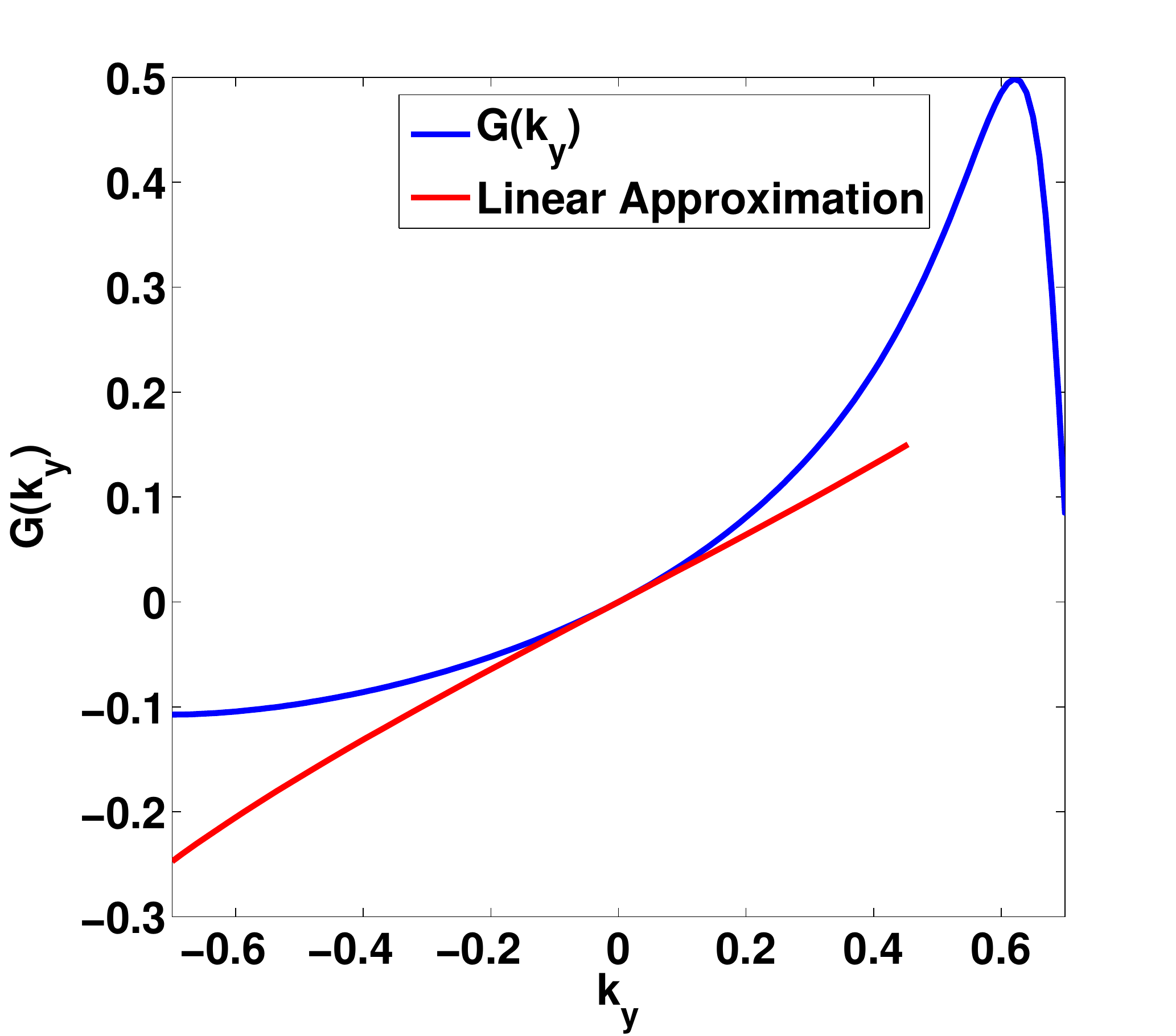}
     }
     \subfloat[\label{subfig-2:dummy}]{%
       \includegraphics[width=0.25\textwidth]{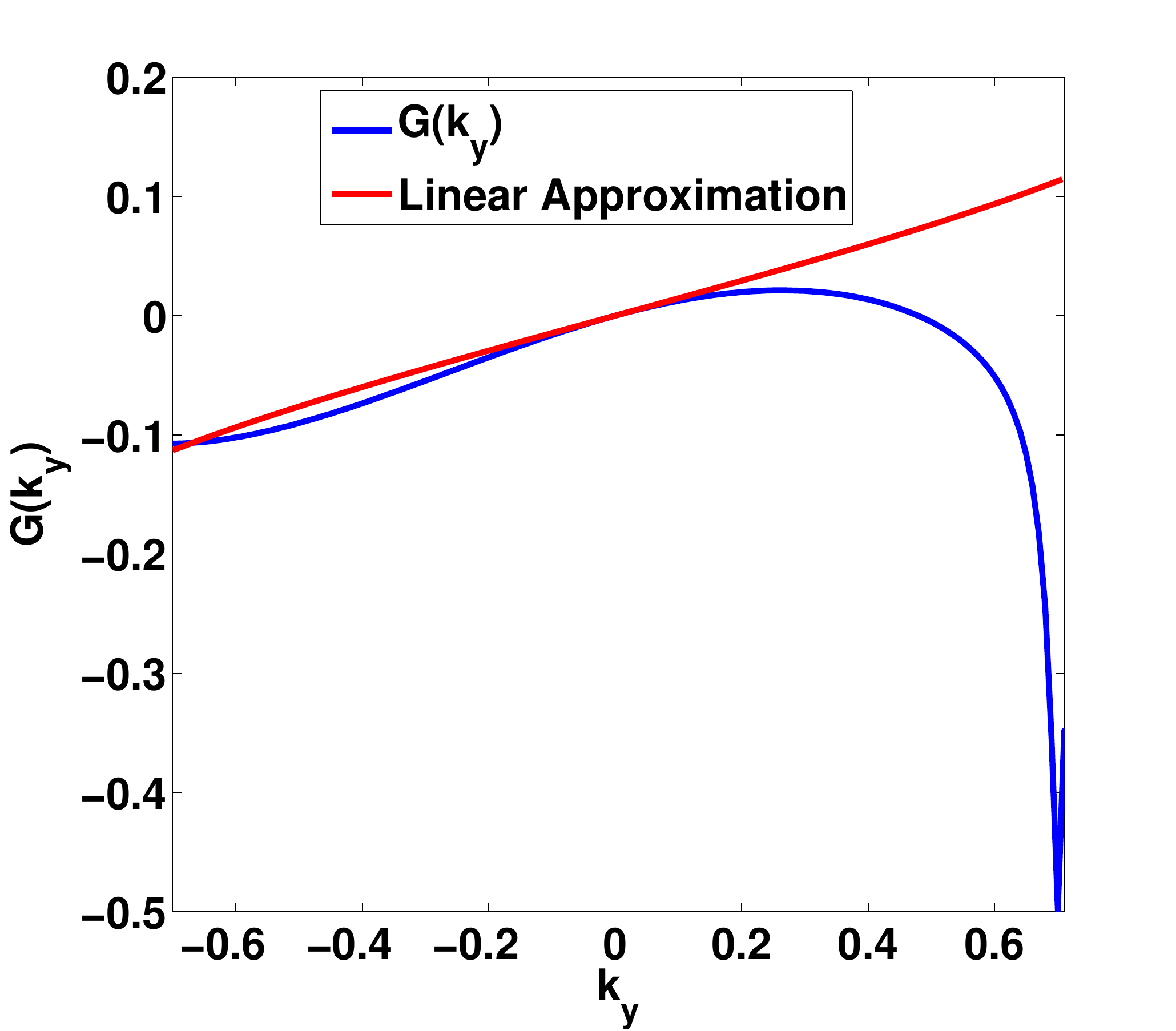}
     }
     \caption{Green's function $G(k_{y})$ of the structure for (a) TE and (b) TM polarization.}
     \label{fig:dummy}
\end{figure}
\indent Here, we design our structure so as to realize the first order derivative of the input field $f(y)$ for the incident angle of $\theta_{1}=45^{\circ}$. For the sake of simplicity, the two semi-infinite dielectric media are taken to be free space, i.e. $n_{1}=1$. The dielectric layer is taken to be silicon (Si) with the refractive index of $n_{2}=3.4$. The required length $L$ for which the dielectric layer acts like a half-wavelength transmission line  will then be obtained as $L=0.9447/k_{0}$.\\
 \indent Figs. 3 (a) and (b) show the associated Green's function $G(k_{y})$ for TE and TM polarizations, respectively,  together with their linear approximations about $k_{y}=0$. It can be seen that both Green's functions can be estimated well by a linear approximations for sufficiently small values of $\triangle k_{y}$, thus, as long as the spatial bandwidth of the input signal $f(y)$ is small. Once the Green's function of the structure is estimated properly by its linear approximation, the reflected field $h(y^{\prime \prime})$ will be nothing except the first-order derivative of the input signal $f(y)$ multiplied by a fixed scale factor.\\
\begin{figure}
\centering
     \subfloat[\label{subfig-1:dummy}]{%
       \includegraphics[width=0.35\textwidth]{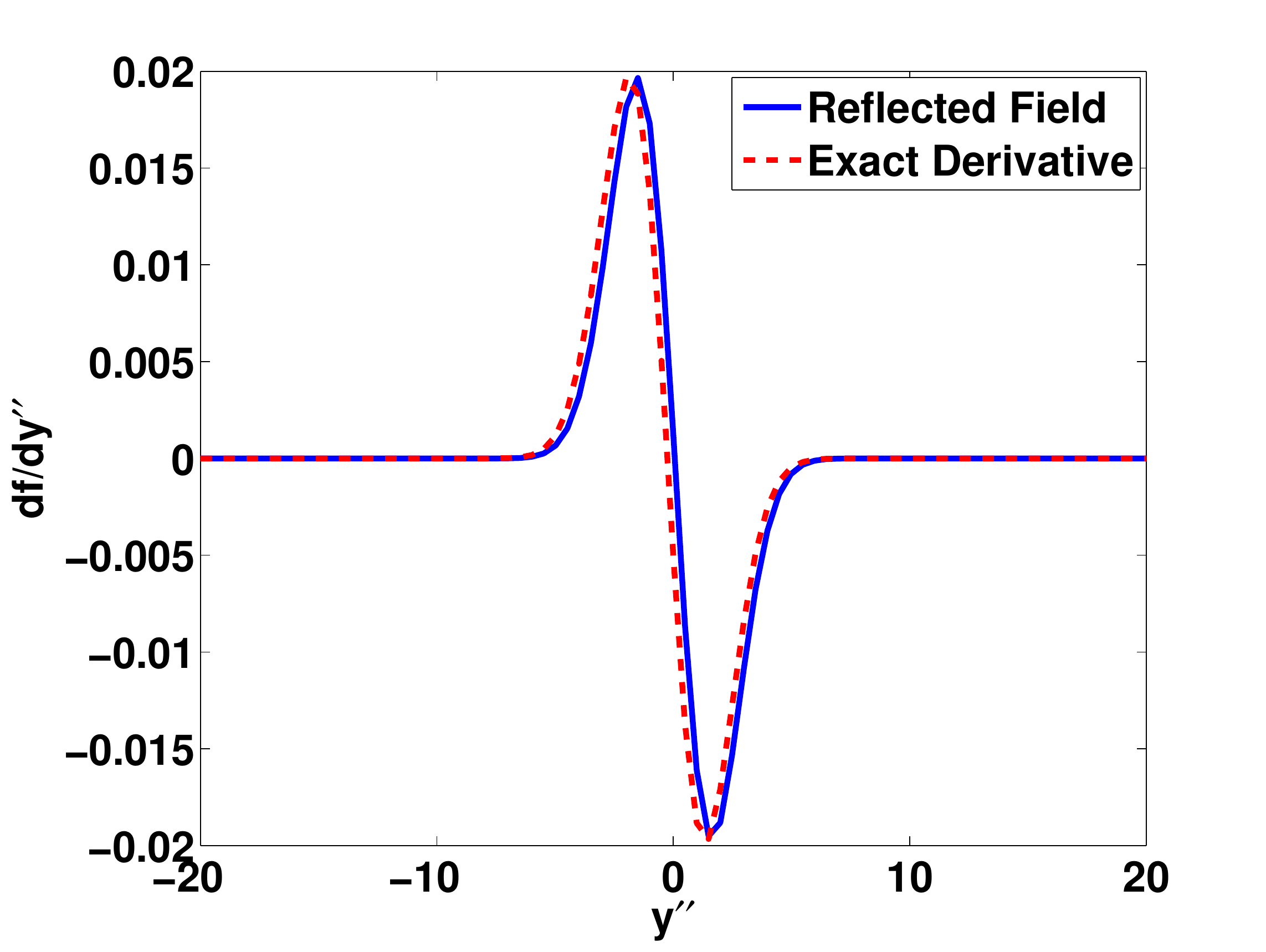}
     }
     \hfill
     \subfloat[\label{subfig-2:dummy}]{%
       \includegraphics[width=0.24\textwidth]{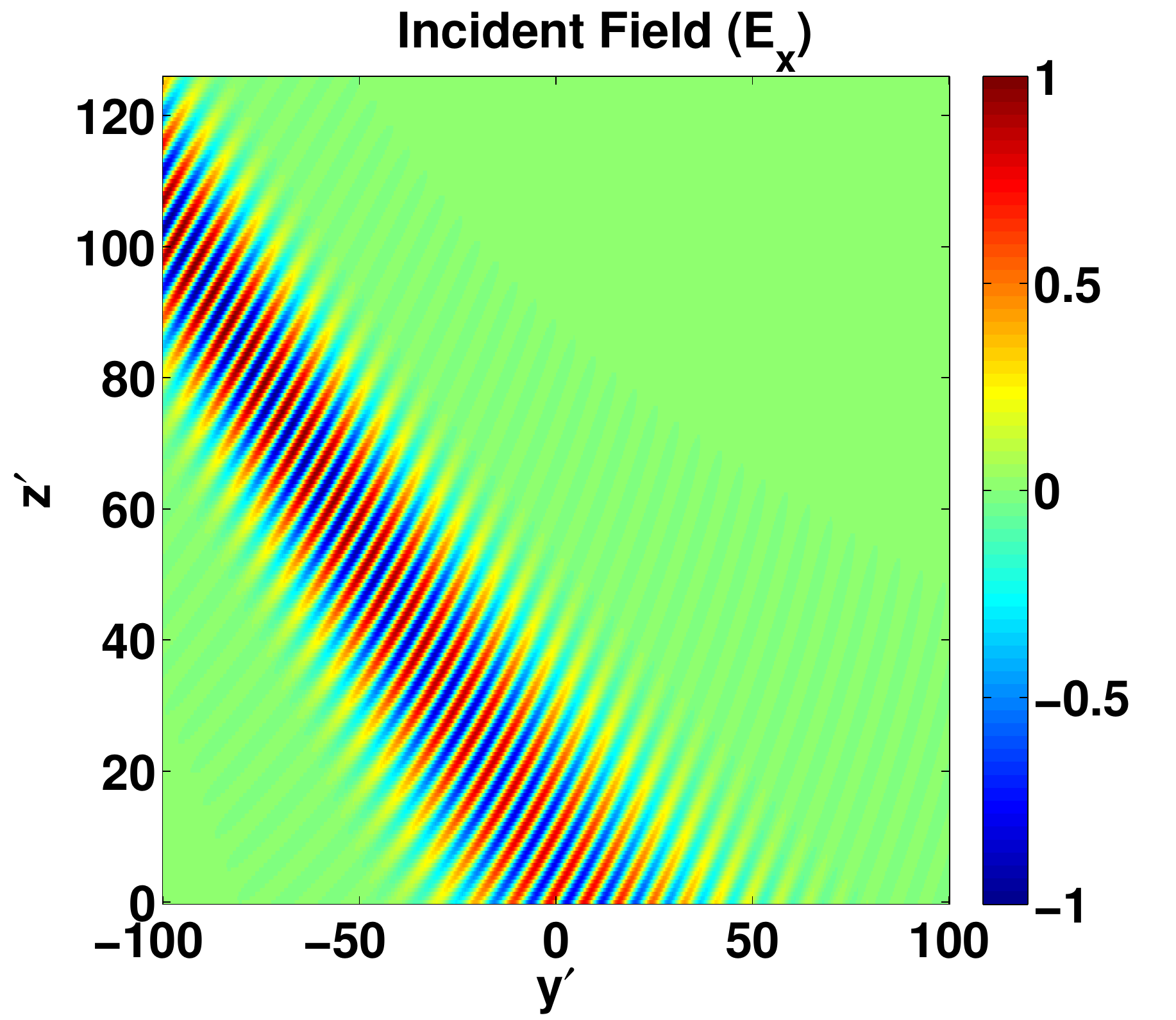}
     }
     \subfloat[\label{subfig-2:dummy}]{%
       \includegraphics[width=0.24\textwidth]{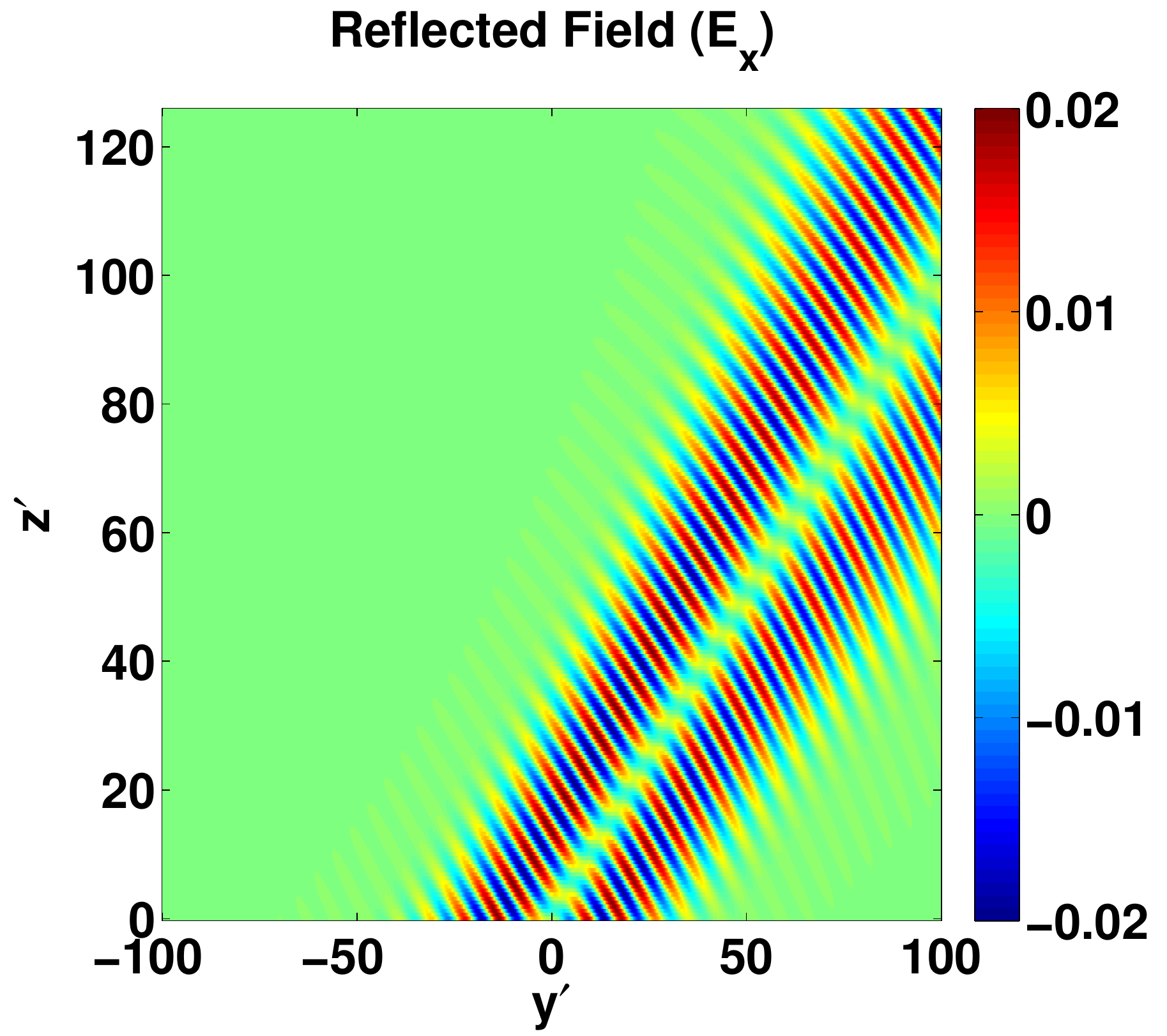}
     }
     \caption{(a) Reflected field together with  scaled calculated first order derivative of input filed (b) TE polarized incident field profile (c) TE polarized reflected field profile.}
     \label{fig:dummy}
\end{figure}
 \indent To inspect the performance of the presented differentiator, a TE polarized Gaussian incident field $f(y)$ with a spatial bandwidth of $W=0.1k_{0}$ (or a beamwidth of $32\lambda_{0}$) is first assumed to be incident on the structure at an angle of $\theta_{1}=45^{\circ}$. Fig. 4(a) illustrates the reflected field $h(y^{\prime \prime})$ together with the scaled calculated first order derivative of the input filed $f(y)$. Te results are in excellent agreement confirming the accurate performance of the differentiator. The field profile of the incident field and reflected field are also shown in Figs. 4(b) and (c).
The results for an incident TM field with the same profile are shown in Fig. 5(a). Similar to the previous case, the results match each other well. The field profile of the incident and reflected fields are also given in Figs. 5(b) and (c), respectively.\\
\indent With respect to our previously reported differentiator in \cite{11}, the proposed half-wavelength dielectric differentiator has the advantage of working properly for both TE and TM polarizations of the incident wave. However, such a differentiator still suffers from dependence on angle of incidence of the input field. To overcome this problem, we turn to a different structure that is based on a half-wavelength plasmonic graphene film, taking the advantage of the tunability of the graphene surface conductivity.\\
   \begin{figure}
\centering
     \subfloat[\label{subfig-1:dummy}]{%
       \includegraphics[width=0.35\textwidth]{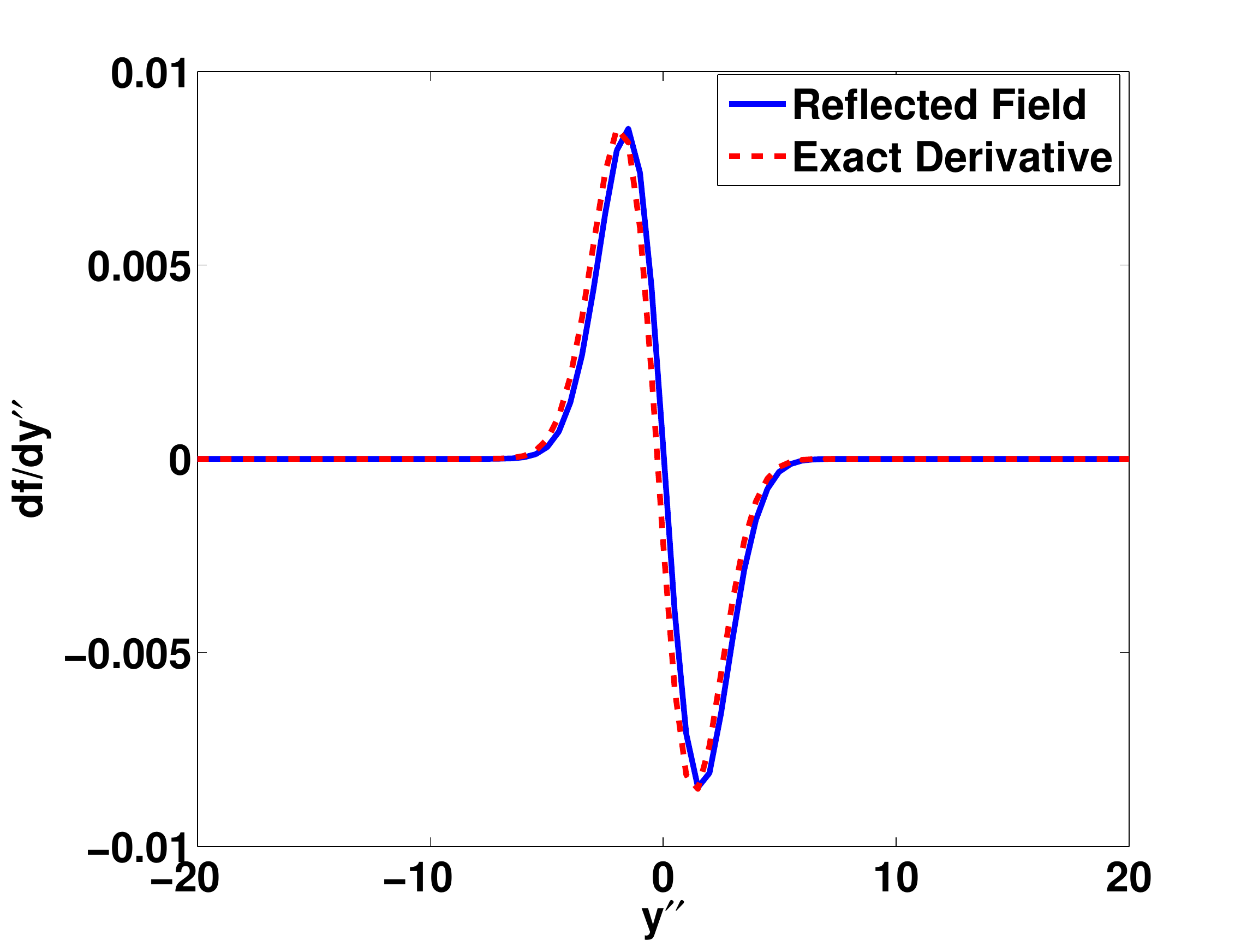}
     }
     \hfill
     \subfloat[\label{subfig-2:dummy}]{%
       \includegraphics[width=0.24\textwidth]{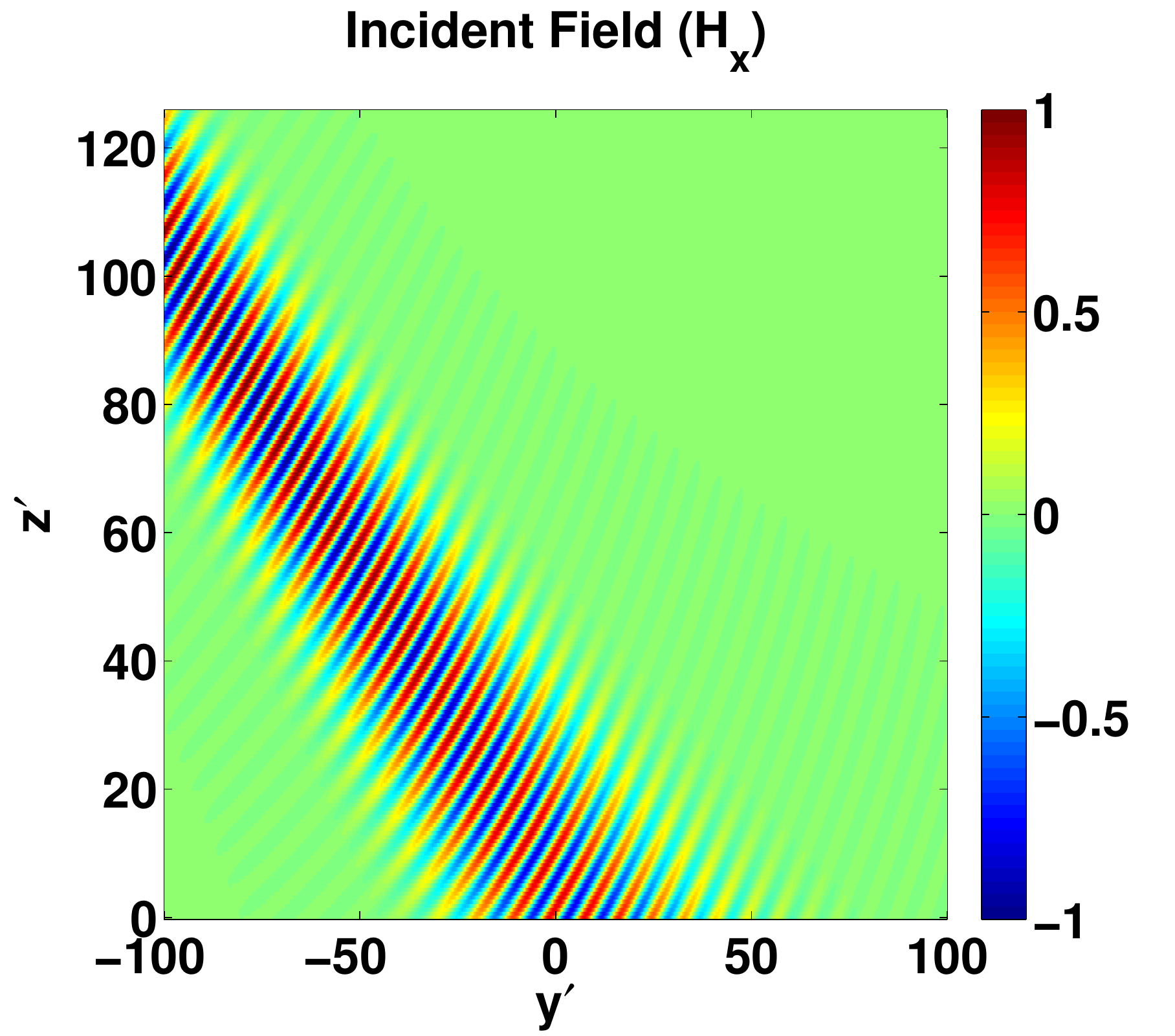}
     }
     \subfloat[\label{subfig-2:dummy}]{%
       \includegraphics[width=0.24\textwidth]{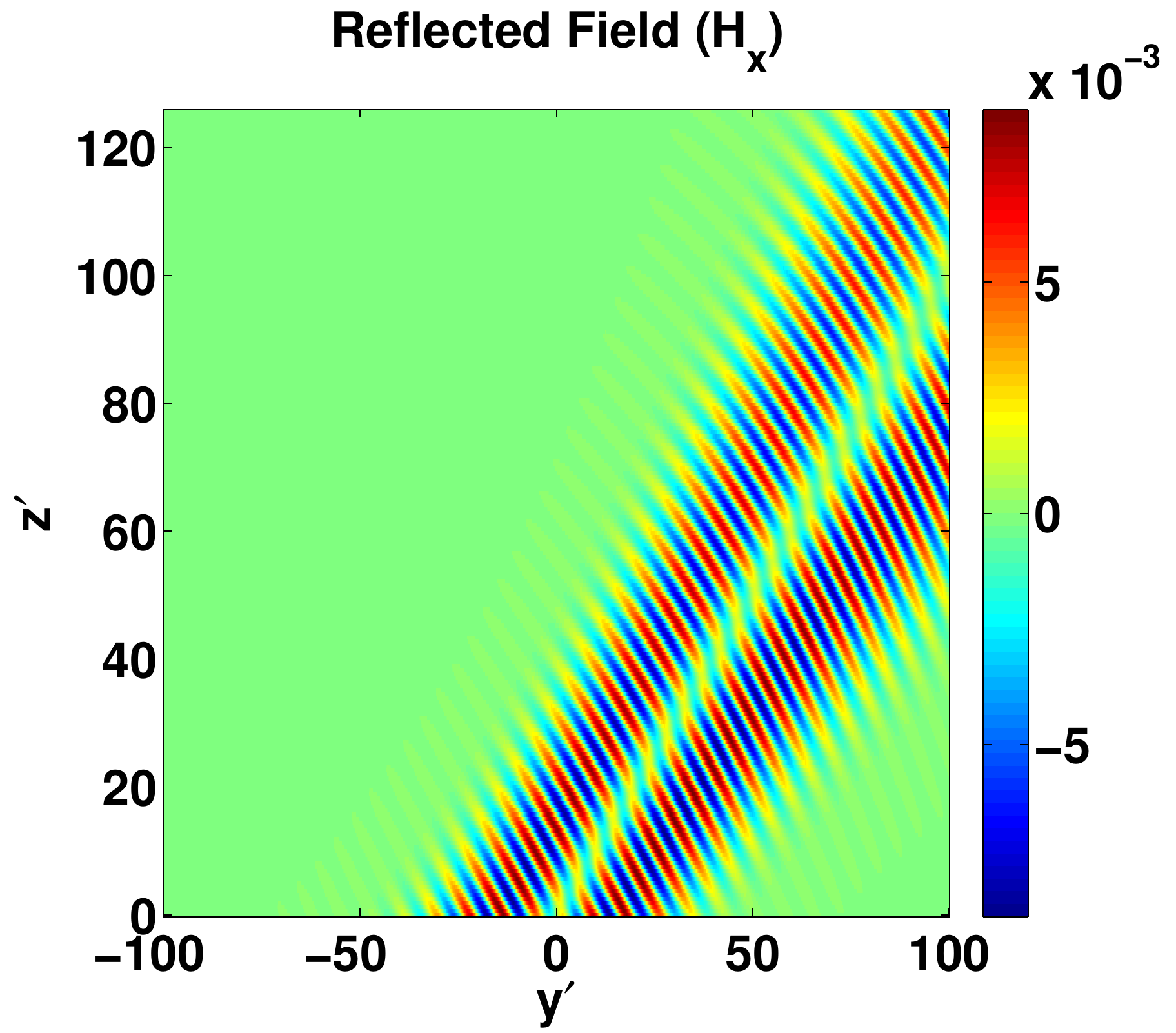}
     }
     \caption{(a) Reflected field together with  scaled calculated first order derivative of input filed (b) TM polarized incident field profile (c) TM polarized reflected field profile.}
     \label{fig:dummy}
\end{figure}
\indent The configuration of the proposed graphene-based differentiator is quite similar to that of the dielectric one shown in Fig. 1. The only difference is that the dielectric media with permittivities $n_{1}$ and $n_{2}$ are replaced by graphene sheets with surface conductivities of $\sigma_{1}$ and $\sigma_{2}$, respectively. The surface conductivity of graphene is expressed in terms of its chemical potential using Kubo formula \cite{13}, and can be easily tuned via a gate voltage acting on it. It is well known that the inductive nature of $\sigma_{1}$ and $\sigma_{2}$ allows the graphene film to support plasmonic waves. The associated propagation constant of such waves can be expressed using a quasi-static approximation as \cite{14}
\begin{equation}\label{3}
 \beta_{p_{1,2}}=\frac{2i\omega\varepsilon_{0}}{\sigma_{1,2}}
\end{equation}
where $\omega=2\pi f$ is angular frequency. Similar to the dielectric case, if the length $L$ is selected such that $\beta_{p_{2}}Lcos(\theta_{2})=\pi$ the graphene film with the surface conductivity $\sigma_{2}$ will act like a half wavelength transmission line, and consequently, the reflection coefficient of the structure becomes zero. As a result, the reflected field will be the first order derivative of the input field, if the input signal $f(y)$ has a narrow spatial bandwidth. Note that by changing the parameter $\sigma_{2}$ via the gate voltage, the incident angle $\theta_{1}$ for which the reflection coefficient gets to be zero, can be tuned. As a consequence, our differentiator is reconfigurable.\\
\indent We first aim to design our structure so that it realizes the operator of first order derivative for an angle of incidence of $\theta_{1}=30^{\circ}$. We assume typical values $\beta_{p_{1}}=1~ \mu m^{-1}$ and $\beta_{p_{2}}=1/1.5~ \mu m^{-1}$ at the operation frequency of $f=3~THz$ \cite{15}. The required length is then $L=7.12~\mu m$. However, for such a structure, the reflection coefficient becomes zero at $\theta_{1}=31^{\circ}$ which is slightly different from our design target. This is because of the anomalous phase shift in the reflection of graphene plasmons that are incident to the  discontinuities between graphene layers with distinct surface conductivities \cite{15}, which renders the criteria $\beta_{p_{2}}Lcos(\theta_{2})=\pi$ inaccurate. Fig. 6 (a) indicates the associated Green's function of the designed structure. The results are derived by solving the integral equation governing the behavior of surface charge density \cite{14}, and by employing a full wave EM solver, namely HFSS. Like the previous case, for narrow spatial band input fields $f(y)$, the Green's function $G(k_{y})$ can be assumed to be linear around $k_{y}=0$, and consequently, the reflection field will be fist order derivative of the input field. \\
\begin{figure}
\centering
     \subfloat[\label{subfig-1:dummy}]{%
       \includegraphics[width=0.24\textwidth]{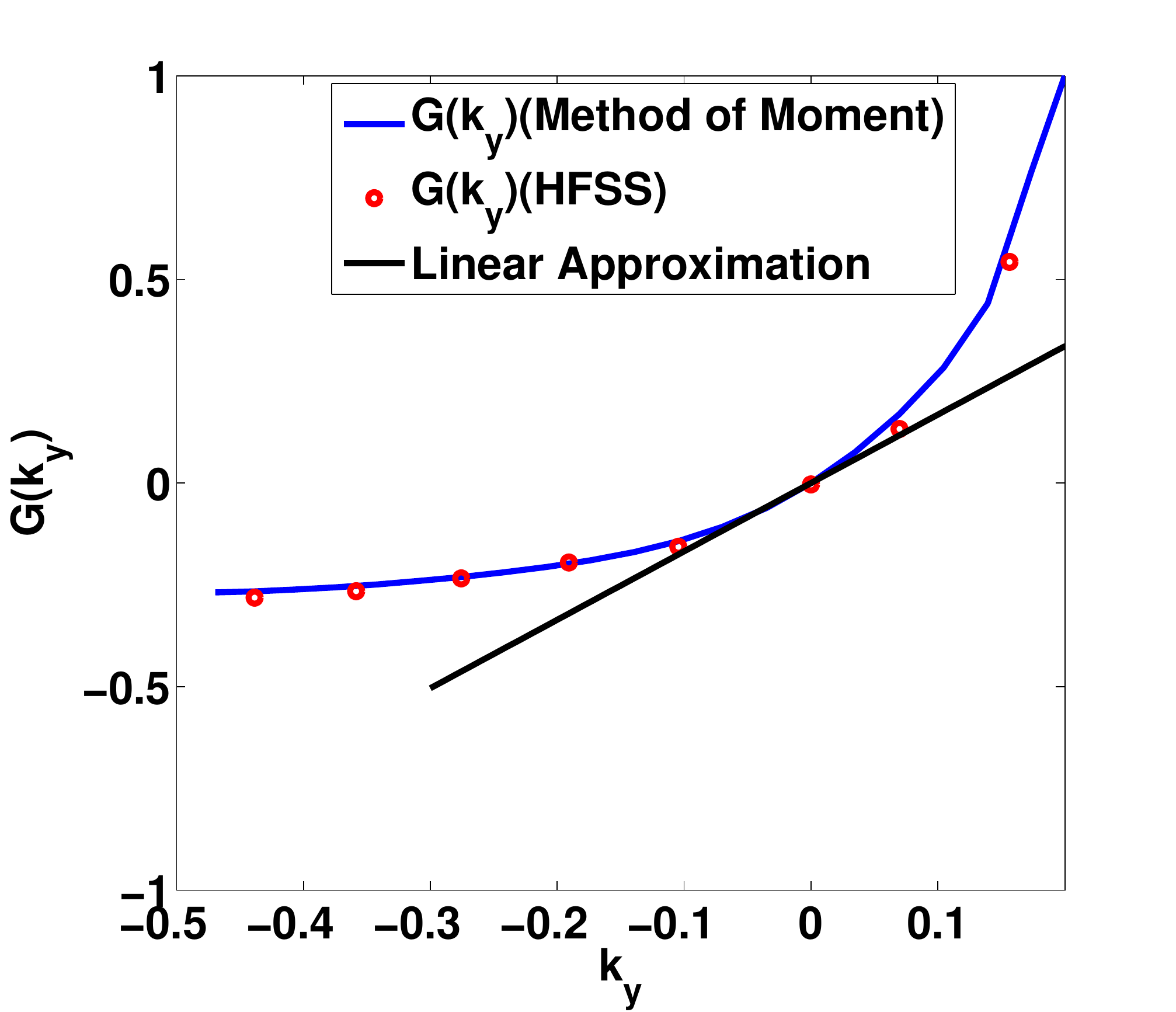}
     }
          \subfloat[\label{subfig-1:dummy}]{%
       \includegraphics[width=0.24\textwidth]{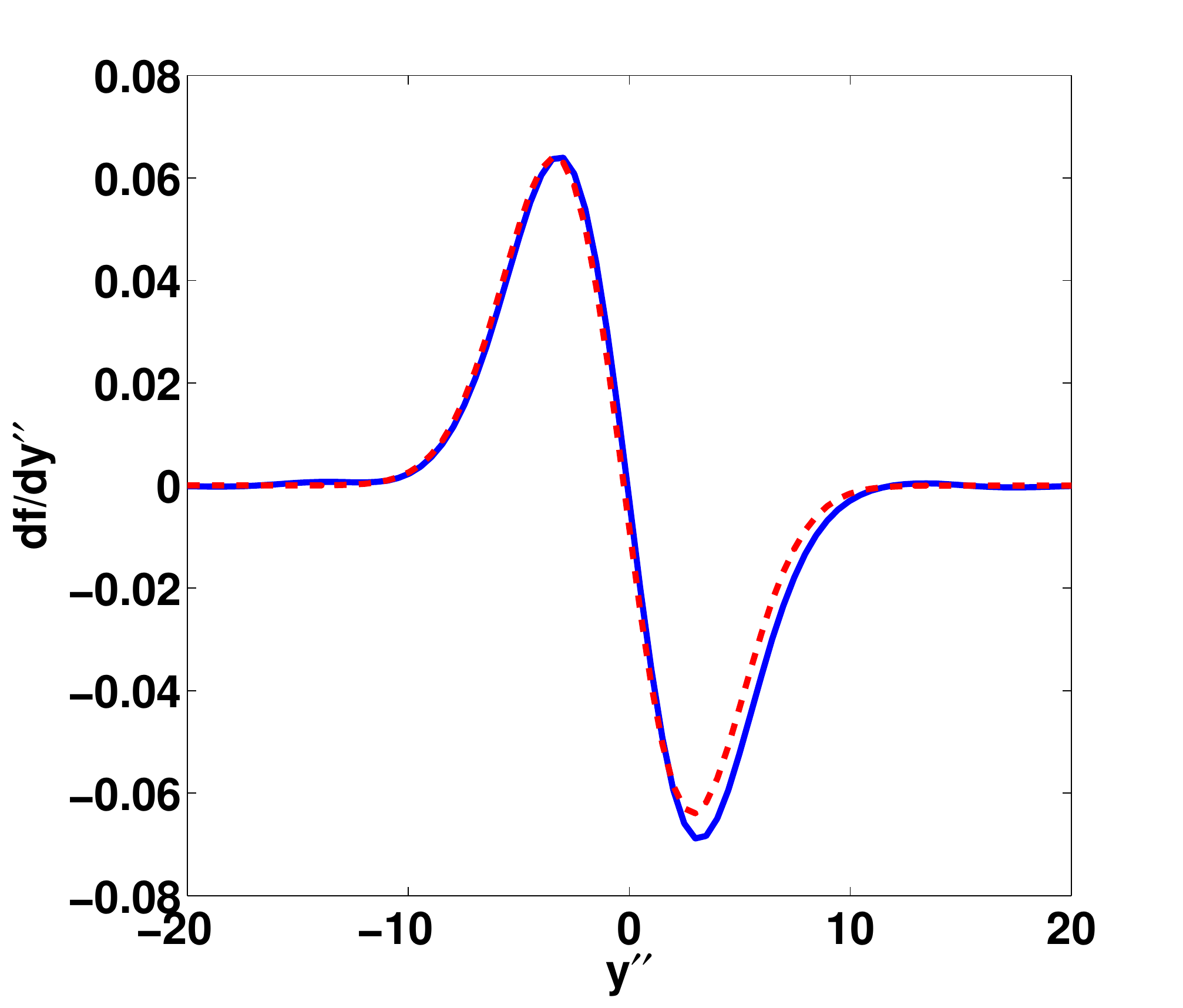}
     }
     \hfill
     \subfloat[\label{subfig-2:dummy}]{%
       \includegraphics[width=0.24\textwidth]{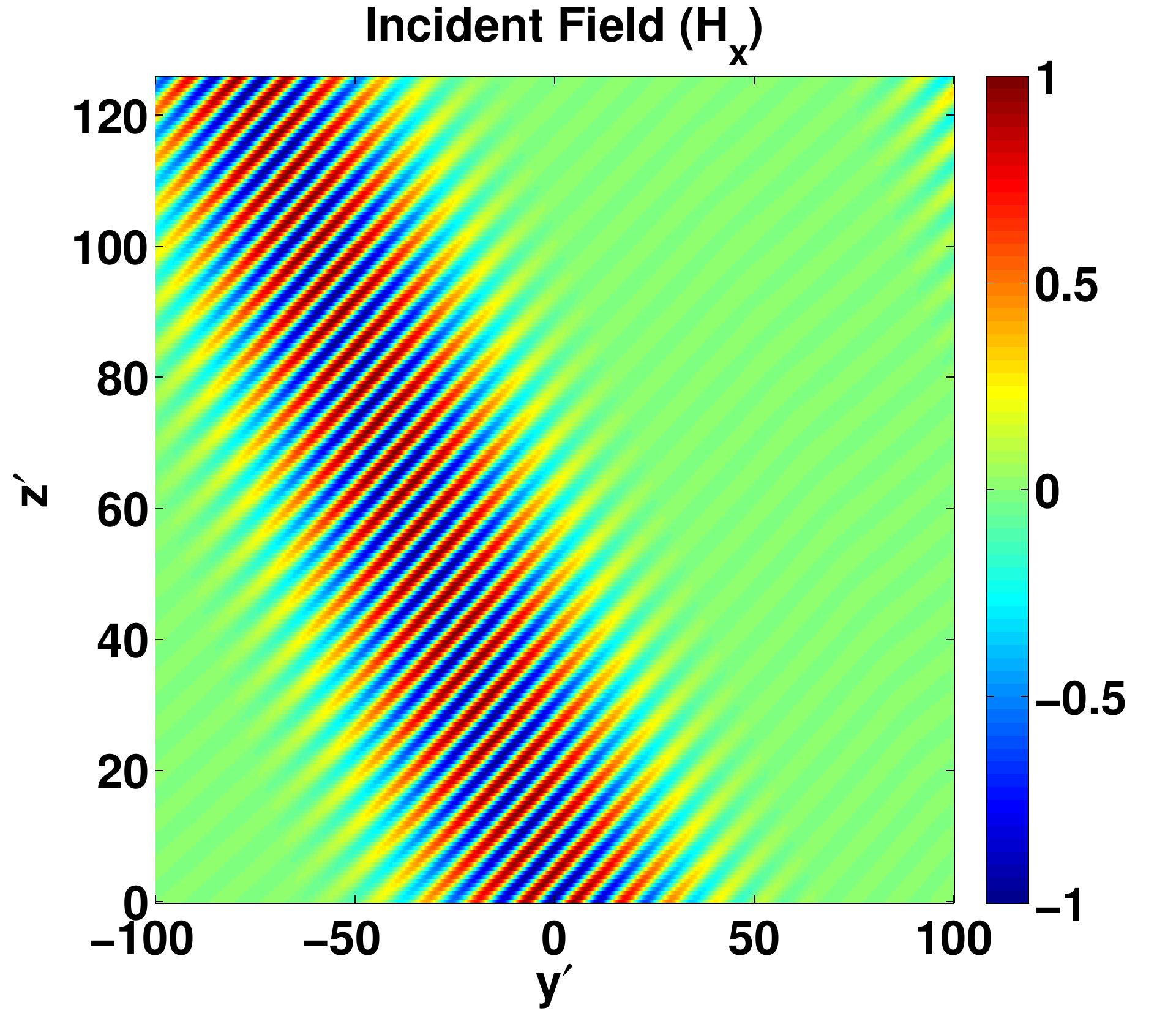}
     }
     \subfloat[\label{subfig-2:dummy}]{%
       \includegraphics[width=0.24\textwidth]{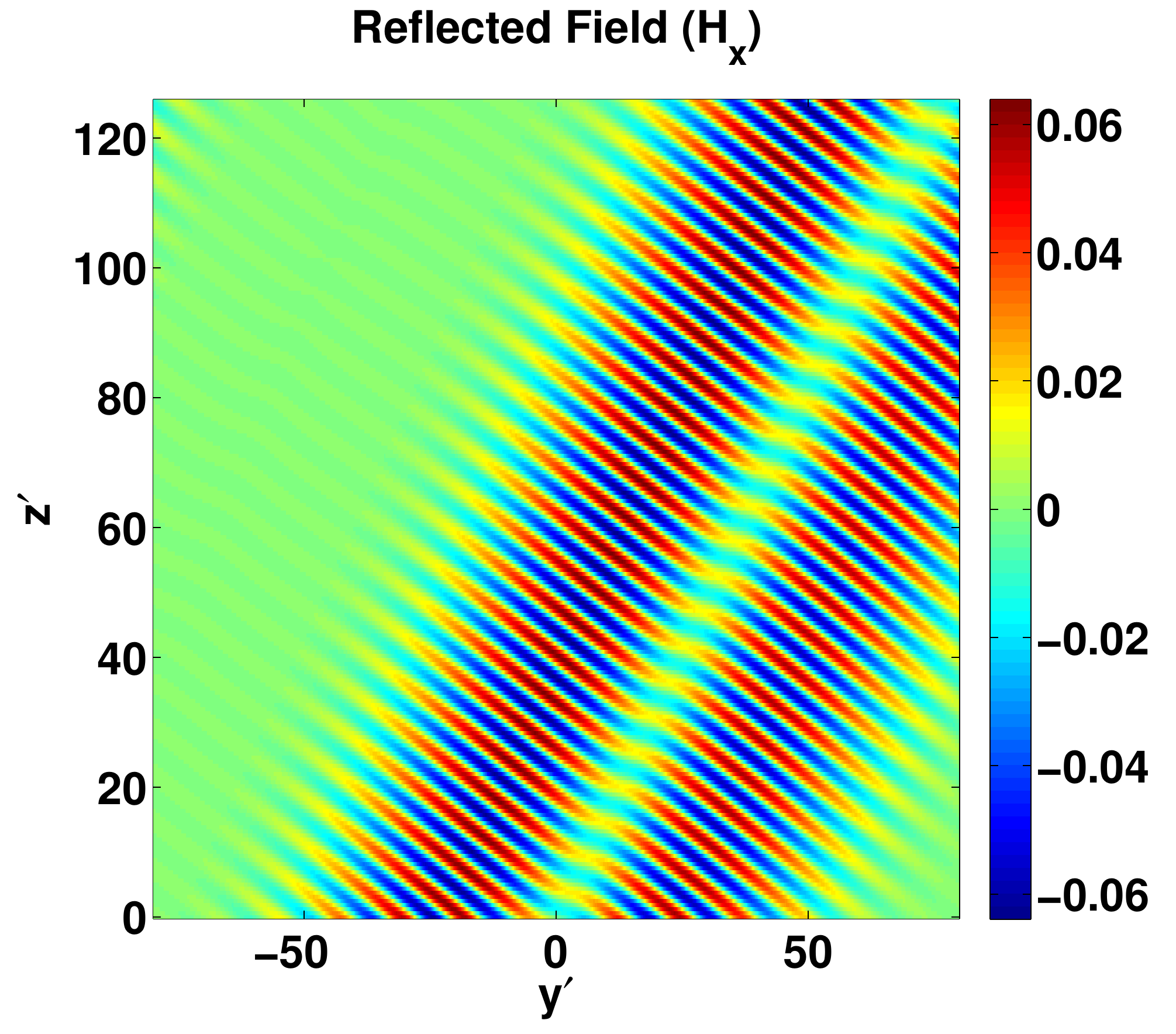}
     }
     \caption{Obtained results of the designed graphene-based differentiator for $\beta_{p_{2}}=1/1.5~ \mu m^{-1}$: (a) Green's function $G(k_{y})$ (b) Reflected field together with  scaled calculated first order derivative (c) Incident field profile (d) Reflected field profile.}
     \label{fig:dummy}
\end{figure}
\indent A Gaussian incident field with a spatial bandwidth of $W=0.05k_{0}$ and incident angle $\theta_{1}=31^{\circ}$  is then considered as the input field. The reflected field together with the scaled calculated first order derivative of the input filed $f(y)$ is plotted in Fig. 6 (b).  The field profile of the incident and reflected fields are also given in Figs. 6 (c) and (d), respectively. The results confirm the appropriate performance of the structure.\\
\indent We then assume the parameter $\sigma_{2}$ is increased such that $\beta_{p_{2}}=1/1.75~ \mu m^{-1}$. The incident angle for which the graphene film with the conductivity $\sigma_{2}$ serves as a half-wavelength transmission line will then be adapted to $\theta_{1}=22^{\circ}$. The associated Green's function $G(k_{y})$, the reflected field $h(y\prime\prime)$, the field profile of the incident and reflected waves are shown in Figs. 7 (a), (b), (c), (d), respectively. These results validate the reconfigurability of the differentiator.

With respect to the aforementioned dielectric-based differentiator, the graphene-based design provides us with two major advantages. The first one is its ability to reconfigure as mentioned. The second one is that it is highly miniaturized as the length $L$ is about one hundred times smaller than the vacuum wavelength. Because of the simplicity of our approach, it is also much more miniaturized than previously reported differentiator employing graphene-based metalines in \cite{12}.
\begin{figure}[!h]
\centering
     \subfloat[\label{subfig-1:dummy}]{%
       \includegraphics[width=0.24\textwidth]{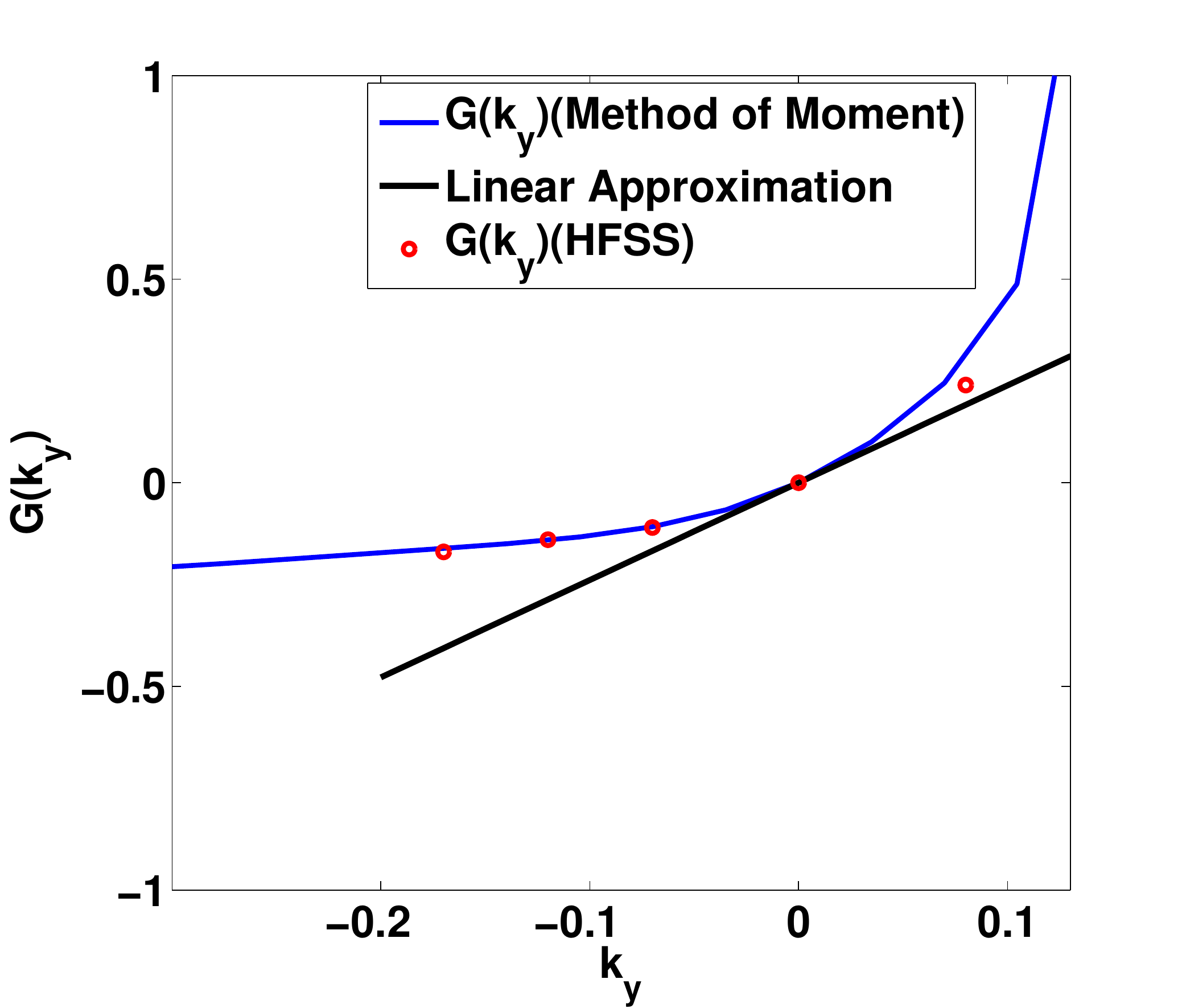}
     }
          \subfloat[\label{subfig-1:dummy}]{%
       \includegraphics[width=0.24\textwidth]{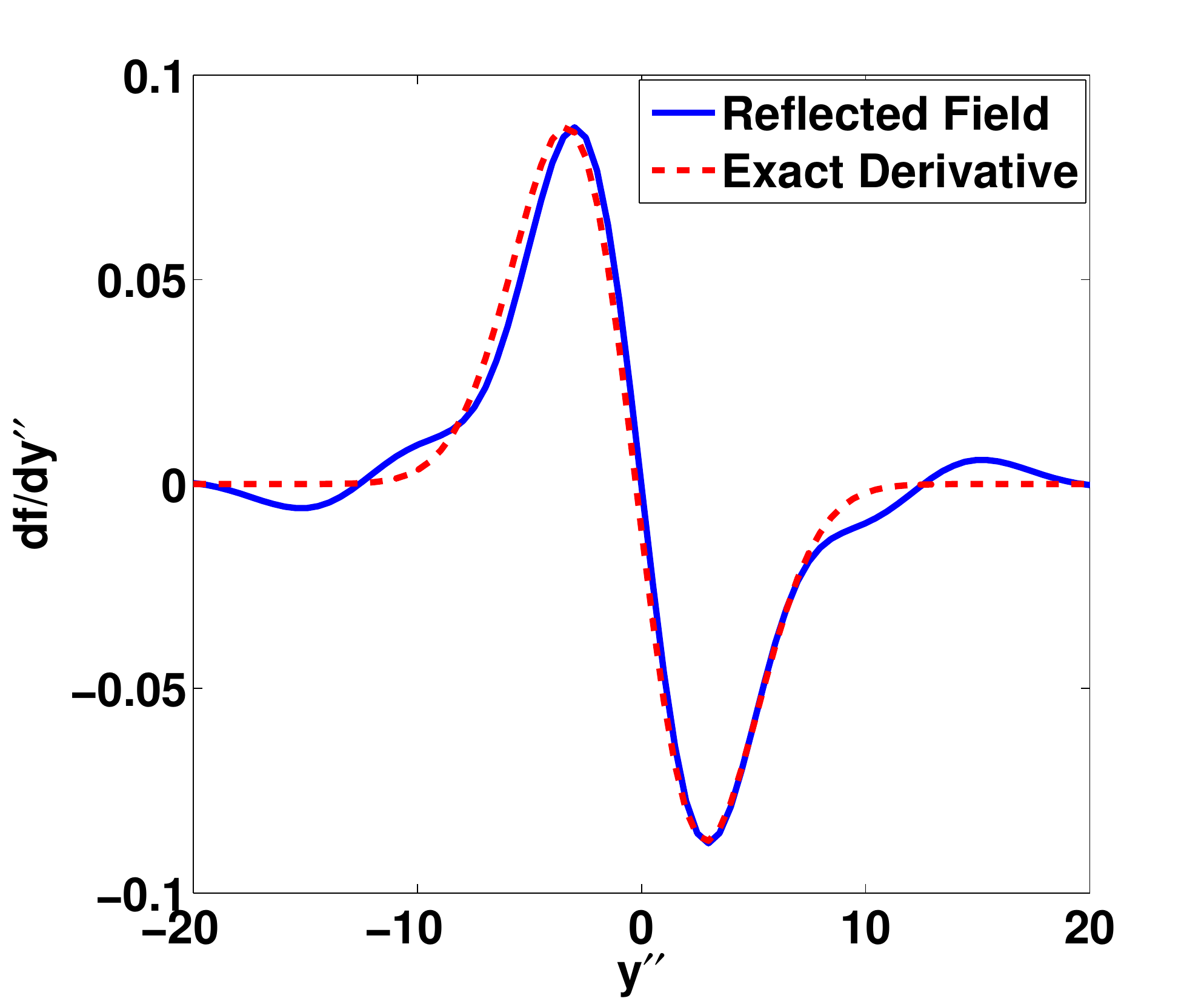}
     }
     \hfill
     \subfloat[\label{subfig-2:dummy}]{%
       \includegraphics[width=0.24\textwidth]{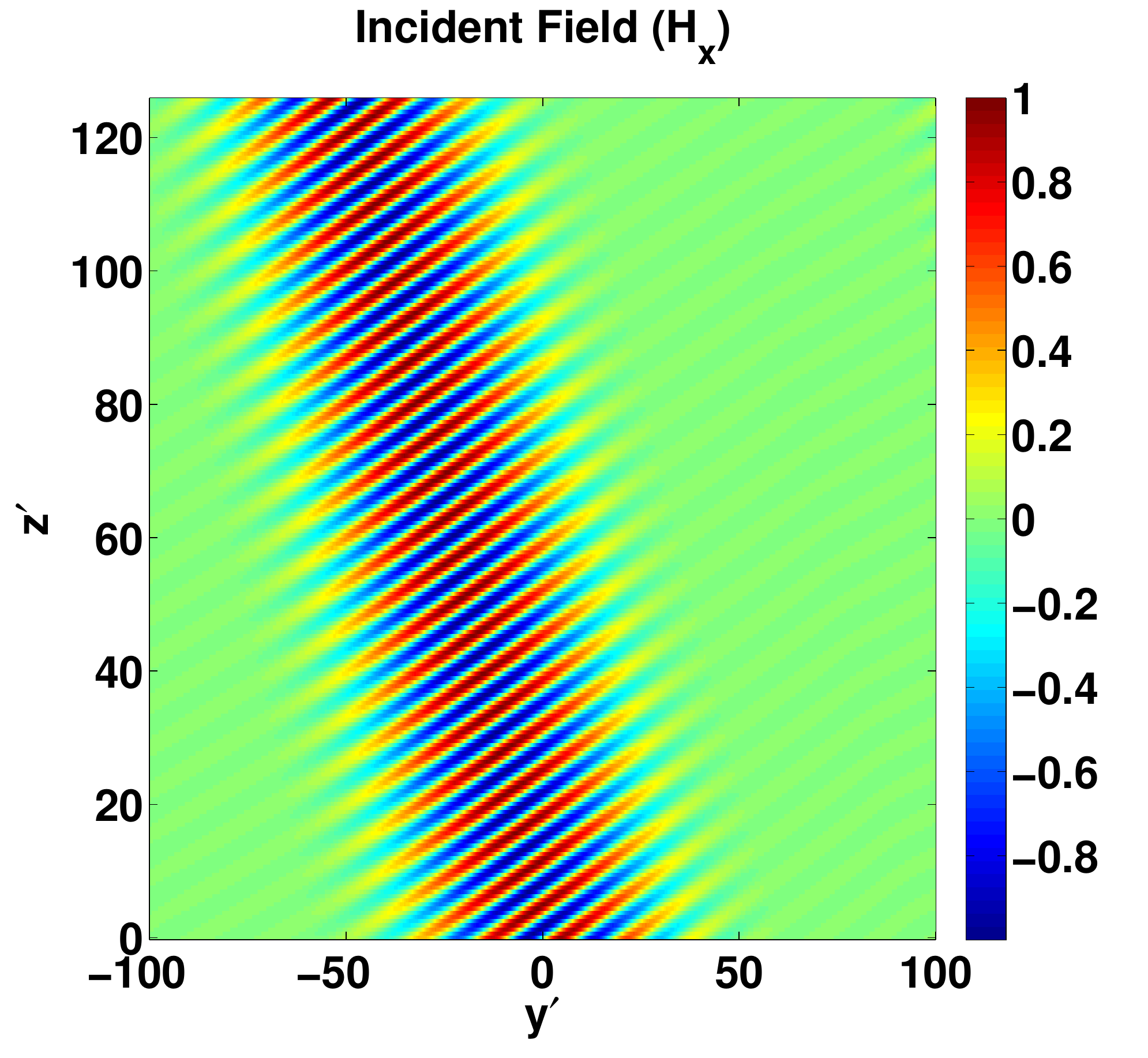}
     }
     \subfloat[\label{subfig-2:dummy}]{%
       \includegraphics[width=0.24\textwidth]{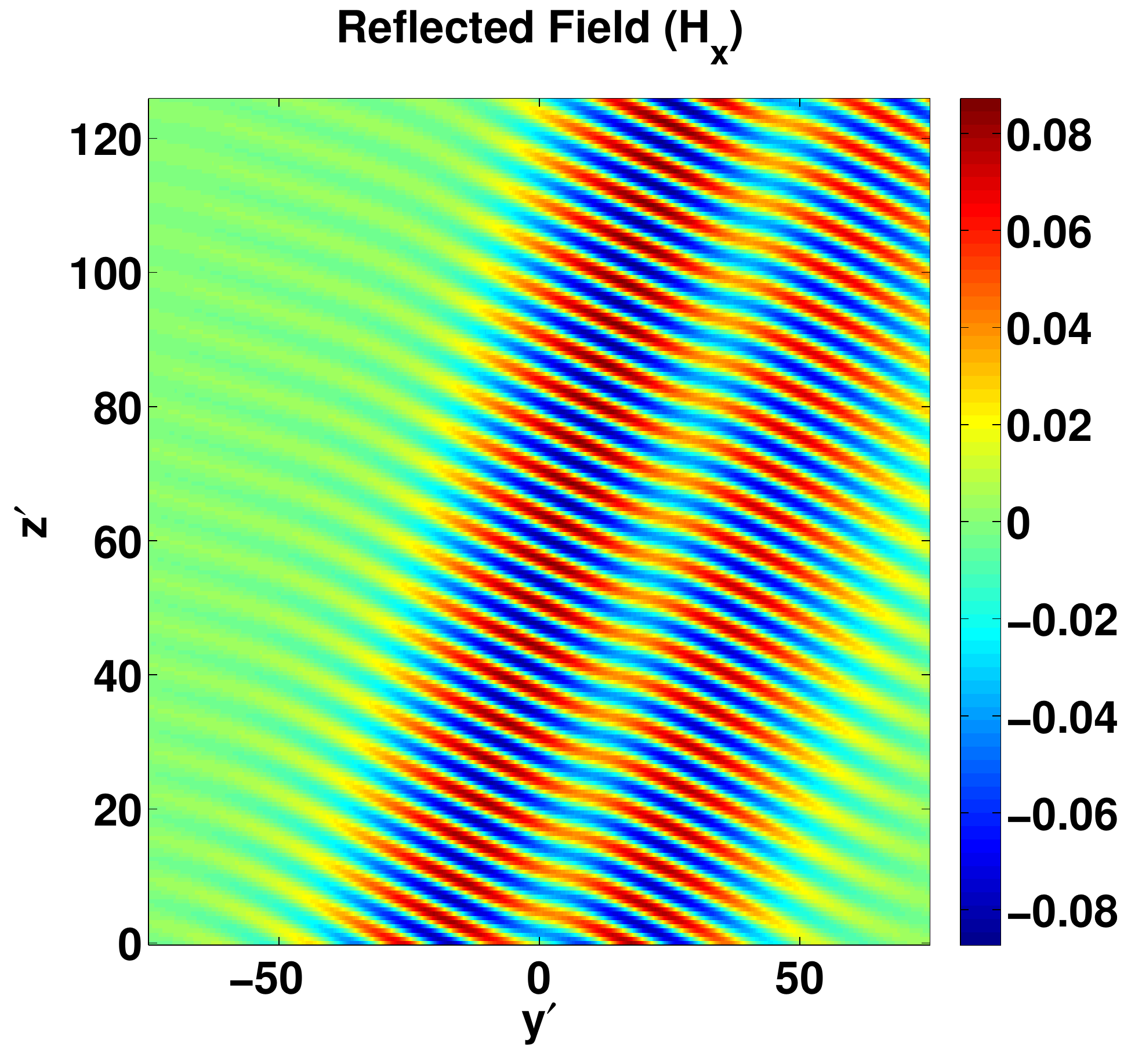}
     }
     \caption{Obtained results of the designed graphene-based differentiator for $\beta_{p_{2}}=1/1.75~ \mu m^{-1}$: (a) Green's function $G(k_{y})$ (b)Reflected field together with  scaled calculated first order derivative (c) Incident field profile (d) Reflected field profile.}
     \label{fig:dummy}
\end{figure}
In summary, we proposed a new approach to realize a differentiator using half-wavelength slabs. We first realized the differentiator making use of a half-wavelength dielectric slab and showed that such a differentiator is polarization independent. Motivated by the good performance of this design, and by exploiting the unique features of graphene,  we then demonstrated a reconfigurable and highly miniaturized differentiator using a half-wavelength graphene film.

    \end{document}